\documentstyle[12pt]{article}

\newcommand{\beq}{\begin{equation}}
\newcommand{\eeq}{\end{equation}}
\newcommand{\ket}[1]{\left| {#1} \right>}

\begin{document}

\title{Active stabilisation, quantum computation and quantum state
synthesis} 

\author{A. M. Steane\\
\small
Department of Physics, Clarendon Laboratory,
Parks Road, Oxford, OX1 3PU, England;\\
\small
Institute for Theoretical Physics, University of California, Santa Barbara}

\date{13 November 1996}

\maketitle

\begin{abstract}
Active stabilisation of a quantum system is the active suppression of
noise (such as decoherence) in the system, without disrupting its unitary 
evolution. Quantum error correction suggests the possibility of achieving
this, but only if the recovery network can suppress more noise than it
introduces. A general method of constructing such networks is proposed,
which gives a substantial improvement over previous fault tolerant designs.
The construction permits quantum error correction to be understood as
essentially quantum state synthesis. An approximate analysis implies
that algorithms involving very many computational steps on
a quantum computer can thus be made possible.
\end{abstract}


All physical systems are subject to noise, arising from
an unavoidable coupling to an
environment which can not be completely analysed or controlled.
Typically, noise is a problem we wish to minimise, and for this purpose
passive and active stabilisation can be applied. By {\em passive} stabilisation
we mean the careful construction and isolation of a system so as to reduce
the noise level to a minimum. By {\em active} stabilisation we mean
the use of detection and feedback to suppress the tendency of a
system to depart from some desired state. An early example 
is the governor in a steam engine, and a common modern example 
is the electronic servo based on a stable voltage reference, a high-performance 
amplifier and a low-noise resistor (see inset to Fig. 2). Passive
stabilisation is insufficient to stabilise any but the most simple 
devices, whereas active stabilisation is a very powerful method and is used 
throughout the natural world, whether in man-made devices or in living 
organisms. 

The above examples of active stabilisation are classical, not quantum
systems, however. Is it possible to apply active stabilisation in quantum 
physics? That is, can the complete 
unitary evolution of the state throughout its Hilbert space be actively 
stabilised against the effects of random noise? 
It has been widely supposed that the answer to this question is `no'. The 
reason is because active stabilisation is based on amplification and 
dissipation. However, a general unknown quantum state cannot be amplified, 
(`no cloning theorem' \cite{clone}), and dissipation will prevent unitary 
evolution, so it would seem impossible to stabilise a quantum state. 

This supposed impossibility was recently called into question by the
concept of quantum error correction (QEC) 
\cite{Shor1,Steane1,CS,Steane2,Laf,Benn,Knill,EM}, which has shown that 
very powerful techniques exist for restoring a quantum 
state which has been affected by noise, and error correction can be applied in 
a manner (dubbed `fault tolerant' \cite{ft,DiVS}) which is itself sufficiently 
insensitive to noise during the corrective process to allow an overall 
stabilisation of a quantum system. 

In this paper I will present a reinterpretation of QEC, showing that
the task of applying QEC in the lab is essentially one of {\em quantum state 
synthesis}, that is, the preparation of a desired (non-trivial) quantum state.
This is more than merely a change of emphasis, since
the proposed method allows a substantial improvement on the best previously
described correction technique. I will then estimate
the degree of passive stability and redundancy needed to enable
a quantum computer to carry out long computations. 

First let us summarise the concept of QEC. 
Suppose we wish to store or communicate an unknown quantum state $\ket{\phi}$
of some quantum system $q$, in the presence of noise. We introduce
an extra system $c$, similar to $q$, in 
an initial known state $\ket{0}$. A unitary `encoding' operation $E$ is 
performed, $\ket{\phi} \otimes \ket{0} \rightarrow E(\ket{\phi} \otimes 
\ket{0}) \equiv \ket{\phi_E}$. The encoded state 
$\ket{\phi_E}$ is transmitted or stored, during which time it is subject to 
noise, $\ket{\phi_E} \rightarrow e_s \ket{\phi_E}$.
Now, such noise will not degrade the
transmitted quantum information, for all error operators $e_s$ such that
\cite{Peres} 
  \beq
E^{\dagger} e_s \ket{\phi_E} = \ket{\phi} \otimes \ket{s}. \label{decode}
  \eeq
Furthermore, the `decoding' operation $E^{\dagger}$ will recover the correct 
quantum state $\ket{\phi}$ after noise processes of the form $\ket{\phi_E} 
\rightarrow \sum_s e_s \ket{\phi_E}$, and even after mixtures of such processes 
\cite{CS,Steane2,Knill,EM,remark}. The latter possibility includes the case of 
entanglement with an unknown environment, which is an important source of 
error. The theory of QEC involves identifying encodings $E$ such
that the set ${\cal S} = \{ e_s \}$ of correctable errors 
includes those {\em most likely} to be produced by the actual noise to which 
the system is subject. 

In the rest of this letter, it will be convenient to consider the case that $q$ 
and $c$ are sets of two-state systems (quantum bits or qubits). Let $k$ be the 
number of qubits in $q$, and $n-k$ be the number of qubits in $c$. The combined 
system will be referred to as $qc$ (also a mnemonic for `quantum channel'). A 
general error of a single qubit can be expressed as a sum \cite{remark} of 
operators taken from the set ${\cal P} = \{ I, \sigma_x, \sigma_z, \sigma_x 
\sigma_z = i \sigma_y \}$, where $I$ is the identity (corresponding to no 
error) and $\sigma_i$ are the Pauli spin operators. A general error of $n$ 
qubits is a sum of tensor products of such single-qubit errors. 

Each type of noise will have a corresponding quantum error correcting 
code (QECC) designed to deal with it. In what follows we will analyse the 
case where the noise involves all the Pauli spin operators, but affects 
different qubits {\em independently}. This is sufficient to cover many 
realistic situations. Let $\epsilon$ be
the order of magnitude of erroneous terms in the density matrix
of $qc$ caused by the noise.
A $t$-error correcting code is defined to be one which can correct 
any error where up to but no more than $t$ qubits out of the $n$ are 
defective. This allows the 
decoding operation to be successful (equation (\ref{decode})) for all the 
terms in the noisy density matrix involving up to $t$ defective qubits. The 
remaining uncorrected part has a relative magnitude of order 
$\epsilon^{(t+1)} C(n,t+1)$, which is very small for $\epsilon n < t$ and $t 
\gg 1$, so the noise is strongly suppressed ($C(n,t)$ is the binomial 
coefficient $n! / t!(n-t)!$). Such powerful QECCs  
were first discovered by Calderbank and Shor \cite{CS} and myself 
\cite{Steane2}; they are related to pioneering work of Shor \cite{Shor1} and 
myself \cite{Steane1}. 

So far we have discussed QEC as it might be applied to a communication 
channel, in which the noise occurs in the channel but is not present in 
the operations $E$ and $E^{\dagger}$. To consider active stabilisation, it 
is necessary to relax this assumption. We will require the {\em 
recovery operator} $R$, 
defined by $R \;e_s \ket{\phi_E} \otimes \ket{0}_a = \ket{\phi_E} \otimes 
\ket{s}_a, \;\; \forall e_s \in {\cal S}$. This is an interaction
between $qc$ and an 
ancilla $a$ consisting of a further $n_a$ qubits introduced
in an initial state $\ket{0}_a$. The recovery corrects the encoded system 
$qc$ without decoding it, carrying the noise into $a$.

An example quantum network to perform recovery is shown in 
Fig. 1. This network is for a $[[n,k,2t+1]] = [[5,1,3]]$ 
single-error correcting code in which a single qubit is encoded into five
\cite{Laf,Benn}. The network can be obtained directly from the {\em 
stabiliser} of the code \cite{Gott,CRSS}. In this case the stabiliser
is, in the notation of \cite{CRSS},
  \beq
{\cal H} = \left( {\cal H}_x \right| \left. {\cal H}_z \right)
= \left( \begin{array}{c} 
11000\\
01100\\
00110\\
00011
\end{array} \right| \left. \begin{array}{c}
00101\\
10010\\
01001\\
10100
\end{array} \right).   \label{Hstab}
  \eeq
The quantum exclusive-or ({\sc xor})
gates shown in Fig. 1 are positioned exactly where the 1's
occur in the two halves of the stabiliser. Their effect is the
transformation 
$e_s \ket{\phi_E} \otimes \ket{0}_a \rightarrow e_s \ket{\phi_E} \otimes 
\ket{s}_a$. The final state of $a$ is measured, to reveal the error
syndrome $s$, which is used to identify the corrective operation
$e_s^{-1}$ to be applied to $qc$ (see 
\cite{Steane2,Laf,DiVS} for details). 

The network in Fig. 1 will not function well in the presence 
of noise, since errors in one qubit can be transported to several others 
by the {\sc xor} operations, and errors in $a$ 
go uncorrected. I now propose to replace this general 
type of recovery network by the type illustrated in Fig. 2. The new network 
requires a larger ancilla, $n_a = 2n$ instead of $n-k$, but has the great 
merit of reducing interactions between $a$ and $qc$ to a minimum, 
and furthermore each qubit in $a$ only interacts with a single qubit in 
$qc$, so single qubit errors in $a$ are only carried onto single qubits in 
$qc$. The latter feature was also achieved in the `fault tolerant' design of 
DiVincenzo and Shor \cite{DiVS}, but their technique involved of the order 
of $(2t+1)(n+k)$ interactions between $a$ and $qc$, whereas the present
method involves just $2n$ interactions; this is important
because each interaction introduces noise into $qc$.

The network of Fig. 2 is constructed as follows. First, the ancilla
is prepared in a superposition of all states satisfying
the parity checks of ${\cal H}$, where each row of ${\cal H}$
is interpreted as a single $2n$-bit parity check.
Next, $a$ and $qc$ interact in such a way that errors in
$qc$ are carried into $a$.
The idea is to store the error syndrome not directly in the state
of $n-k$ qubits of $a$ as before, but in the value of the $n-k$
parity checks on $a$ given by ${\cal H}$. It is simple to show
that any error $e_s$ (acting on $qc$ before the recovery network
is applied) will result, after the action of the network,
in a final state of $a$ which passes or fails these parity checks
in the correct way to yield the error syndrome. Furthermore,
the $2n$ qubits of $a$ were prepared in an equally weighted
superposition of $2^{2n - (n-k)}$ orthogonal product states, in
a Hilbert space of $2^{2n}$ dimensions, so there is sufficient
Hilbert space remaining for $a$ to store $n-k$ qubits
of quantum information, which is just enough to store the error
syndrome. Therefore, no further information passes from the
$qc$ to $a$, and there is no problem of extraneous
entanglement left between the two systems after their interaction.

We have now reduced the problem of QEC
essentially to one of quantum state synthesis, since the 
only non-trivial part of the recovery network is that involved
in preparing the ancilla state. Comparing with active stabilisation
of classical systems (Fig. 2), we see that the amplifier is
replaced by the use of redundant storage of the quantum information,
the stable reference is replaced by a precise quantum state synthesis,
and the dissipative feedback resistor is replaced by dissipative
measurements on the ancilla (or, if preferred, by a unitary network
to interpret the syndrome, followed by dissipative re-preparation
of $a$). 

The preparation of $a$ can be accomplished in the presence of 
noise by adopting the ideas of purification \cite{QPA} and 
fault-tolerant quantum computing \cite{ft}. The ancilla is prepared 
and then tested before it is allowed to interact with $qc$. If it 
fails a test, the preparation is started over again. Only once a 
`good' ancilla state is obtained is the rest of $R$ applied. The tests 
could consist of measurements of the parity checks which $a$
should satisfy, whether in the $x$ or $z$ bases, performed by {\sc xor}s 
onto an additional qubit introduced for the purpose. Of course noise will 
cause some of these tests to give erroneous results, but random errors
in $a$ are unlikely to go undetected, except those 
occuring during or after the last test operation on each qubit in $a$.
Since there are $n_a$ qubits, there are $n_a$ such 
opportunities for error. These errors will be taken into account in the 
analysis to follow, the most questionable assumption being that they are 
distributed independently amongst the qubits of $a$. 

Noise will also take place during the interaction of $qc$ and $a$, and at 
other times, resulting in an inappropriate syndrome in $a$.
To handle this,
the whole syndrome generation procedure (preparation, interaction, measurement)
is repeated $r$ times (following Shor \cite{ft}).
Together the $r$ syndromes are used 
to deduce the corrective operation most likely to be appropriate for $qc$ 
after the $r$th syndrome was generated. 
It will be assumed that the
probability this final corrective operation is nevertheless wrong
is equal to the probability that more than half of the $r$ syndromes were 
wrong. Note that the $O(r)$ correct syndromes
will not necessarily all be the same; each one identifies the errors in
$qc$ at the time it was generated.

The method of Fig. 2 becomes particularly elegant when the encoding uses 
a $[[n,k=2k_c - n,d]]$ QECC based on a $[n,k_c,d]$ classical weakly 
self-dual code (ie the codes first discussed in \cite{CS,Steane2}). This 
is shown in Fig. 3. For these codes, the correction of $\sigma_x$ and 
$\sigma_z$ errors can take place separately, using an ancilla of $n$ 
qubits twice. Remarkably, the preparation of $a$ is identical to 
first preparing it in the encoded zero state of $qc$, $\ket{0_E}$, and then 
carrying out a Hadamard transform! Fig. 3 shows two separately encoded 
qubits interacting by a quantum gate, which we take to be one elementary 
step in a longer quantum computation, followed by a recovery operation 
applied first to one qubit, then to the other. Not shown is the testing 
of $a$, nor the repetition of the syndrome generation. 

To estimate the effects of noise in the whole network shown in Fig. 3,
a simple analysis using classical error probabilities will suffice. This
relies on the more thorough treatment of Knill and Laflamme
\cite{Knill,Conc} who have shown that such an analysis is sufficient
for approximate purposes as long as the noise does not
have certain pathological features.

Let $\alpha$ be the probability that the syndrome obtained in any one cycle 
of syndrome generation does not correctly indicate the errors in $qc$. 
During each cycle there are $n_a = n$ opportunities for the last tests on 
$a$ to leave errors in $a$, $n$ {\sc xor} gates between $qc$ and $a$, and 
$n$ measurements on $a$, making $3n$ opportunities in all for errors in $a$, 
so $1 - \alpha = [(1-\gamma)(1-\epsilon)^n]^{3n}$, giving $\alpha = 3 n 
(\gamma + n \epsilon)$ for $\gamma, \epsilon, \alpha \ll 1$,
where $\gamma$ is the probability of gate or measurement
failure and $\epsilon$ is the probability per time-step
that a freely evolving qubit defects. A failed gate
causes all qubits involved in the gate to become defective; a failed
measurement yields an arbitrary result. The
$\epsilon$ term allows one
time-step per gate, that is, these gates are {\em not} performed in parallel.
The probability that more than half the $r$ syndromes are wrong is 
  \beq
P_1 \simeq \sum_{i=(r+1)/2}^r C(r,i) \alpha^i.   \label{P1}
  \eeq

How many errors will accumulate in $qc$?
The $n$ {\sc xor}s between $qc$ and $a$ during each cycle 
carry all the errors already in $a$ into $qc$, and add a further $n$ 
opportunities for error, making $r 2n$ in all. The original elementary
step in the computation involves $O(n)$ logic gates, using fault-tolerant
computation \cite{ft}. Finally we must add a further $r 2n$ error opportunities
which each half (either $\sigma_x$ or $\sigma_z$ correction) 
of the recovery network causes for the other half. Given that $qc$
is encoded with a $t$-error correcting code, the probability that more
errors accumulate in $qc$ than can be corrected is
  \beq
P_2 \simeq \sum_{i=t+1}^{n(4r +O(1))} C\left(n\left[4r+O(1)\right],i\right)
(\gamma + n\epsilon)^i.
 \label{P2}
  \eeq

The probability that the whole computational step fails is 
$p = 4(P_1 + P_2)$, assuming the two encoded qubits can be recovered
in parallel. Solving this equation for $p$ as a function
of $\gamma$, for given $n,t$, taking $\epsilon = \gamma/10 n$
(ie the noise is dominated by gate and measurement errors),
and choosing $r$ large enough to make
$P_1 < P_2$, yields $p = O((nr\gamma)^{(t+1)})$. The solution
is plotted in Fig. 4 for various QECCs
taken from \cite{SQECC}. The main result is that $\gamma$ must be reduced
to a level less than $\sim 10^{-4}$ before active stabilisation can
work, but below this break-even point the stabilisation is very powerful,
allowing for example $p=10^{-12}$ with $\gamma$ of order $10^{-5}$. 
For a computation involving
$S$ elementary steps, it is only necessary to reduce $p$ to less than $1/S$
for the stabilisation to be deemed sufficient, since then a repetition of the 
whole computation can be used to enhance the chances of getting a correct 
result. Thus $\gamma \simeq 10^{-5}$ would permit a quantum computation 
involving $10^{12}$ steps, which is probably impossible to achieve with passive 
stabilisation alone. 

In conclusion, active stabilisation of a quantum system 
is possible through the use of quantum error correction, 
as long as the recovery network removes more noise than it introduces.
This has been achieved by minimising the interaction 
between the ancilla and the system to be stabilised, 
using only $2n$ quantum {\sc xor} gates, which also prevents errors 
propagating from one qubit to many. The recovery network consists primarily 
in the preparation of the ancilla state, and the syndrome information is 
stored in a subtle form, in the values of parity checks over the ancilla 
qubits. A heuristic analysis of the effects of noise on the whole
recovery network suggests that long quantum computations can thus be
made possible. The method described is a great improvement on the 
previously reported `fault-tolerant' error correctors, and allows
QEC to be reinterpreted as active stabilisation 
based on repeated quantum state synthesis. Future work must analyse in
more detail this state synthesis (ancilla preparation) in the presence
of noise.

I would like to acknowledge helpful discussions with P. Shor
and J. Preskill. This research was supported in part by the National
Science Foundation under Grant No. PHY94-07194.
The author is supported by the Royal Society.

\newpage

\begin{figure}
\caption{Recovery network for a $[[5,1,3]]$ code, showing how the network 
may be built directly from the stabiliser matrix, but yielding a design 
which functions poorly in the presence of noise. The symbol $\bigcirc 
\hspace{-1.9ex} \scriptstyle H$ means the Hadamard transform
$\left| 0 \right> \rightarrow
(\left| 0 \right> + \left| 1 \right>)/\protect\sqrt{2}$,
$\left| 1 \right> \rightarrow
(\left| 0 \right> - \left| 1 \right>)/\protect\sqrt{2}$.
The rectangular box represents a measurement 
of the ancilla bits; the vertical arrow signifies a final corrective 
operation which depends on this measurement result.} 
  \label{fig1}
\end{figure}

\begin{figure}
\caption{Proposed design for the recovery network, illustrated for the
same code as Fig. 1 (cf equation (\protect\ref{Hstab})). The
interaction between the encoded system and the ancilla is now minimal,
and prevents unnecessary propagation of errors.
Inset: the elements of a classical active stabilisation servo.}
\label{fig2}
\end{figure}

\begin{figure}
\caption{A complete computational step, including stabilisation, for
two encoded qubits. In this diagram each horizontal line is a single
encoded qubit, ie $n$ physical qubits, and an open circle at the
beginning of a line means the preparation of the encoded 
zero state $\ket{0_E}$. The vertical arrow represents the corrective
operation $\sigma_x$ or $\sigma_z$ carried out on one or more
qubits.}
\label{fig3}
\end{figure} 

\begin{figure}
\caption{Solution of $p=4(P_1 + P_2)$; $\epsilon = \gamma/10 n$
(cf equations (\protect\ref{P1}) and (\protect\ref{P2})), as a function of 
$\gamma$, for various quantum codes which support fault tolerant 
computation. The kinks in the curves occur when $r$ must be increased in 
order to keep $P_1 <p$; the values of $r$ are in the range 3 to 15. The 
codes are identified by the notation $[[n,k,2t+1]]$. The dashed line 
is for a higher rate code which probably will support universal fault 
tolerant computation, but this has yet to be proved.}
\label{fig4}
\end{figure}


\begin{thebibliography}{99}
\bibitem{remark} To save space, this summary of QEC is written as if only 
noise causing unitary rotations within the Hilbert space of $qc$ is under 
consideration. However, codes which correct the complete set ${\cal P}$ 
will also correct errors which entangle $qc$ with its environment, so in 
fact the discussion is more general, and includes noise processes such as 
relaxation and decoherence (see refs 5,8,9).
\end{thebibliography}
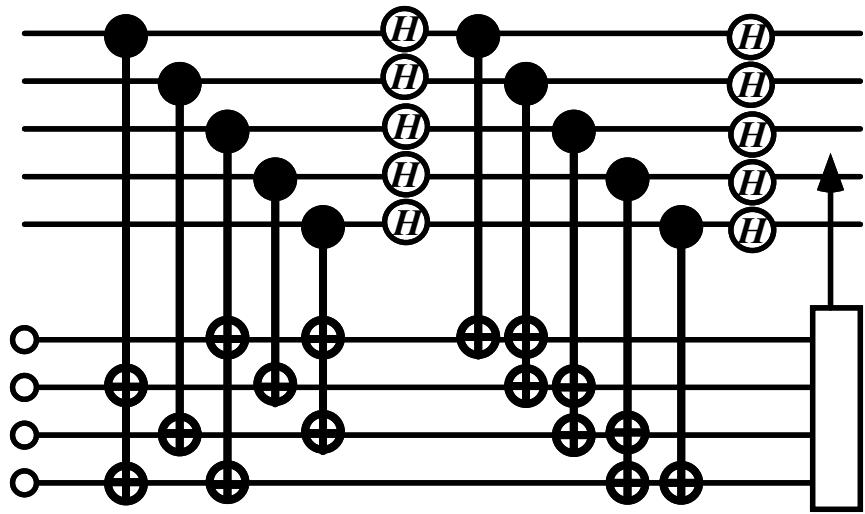
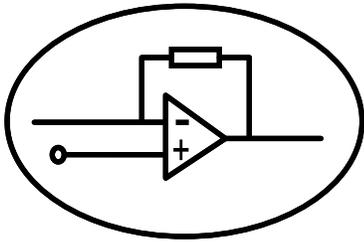
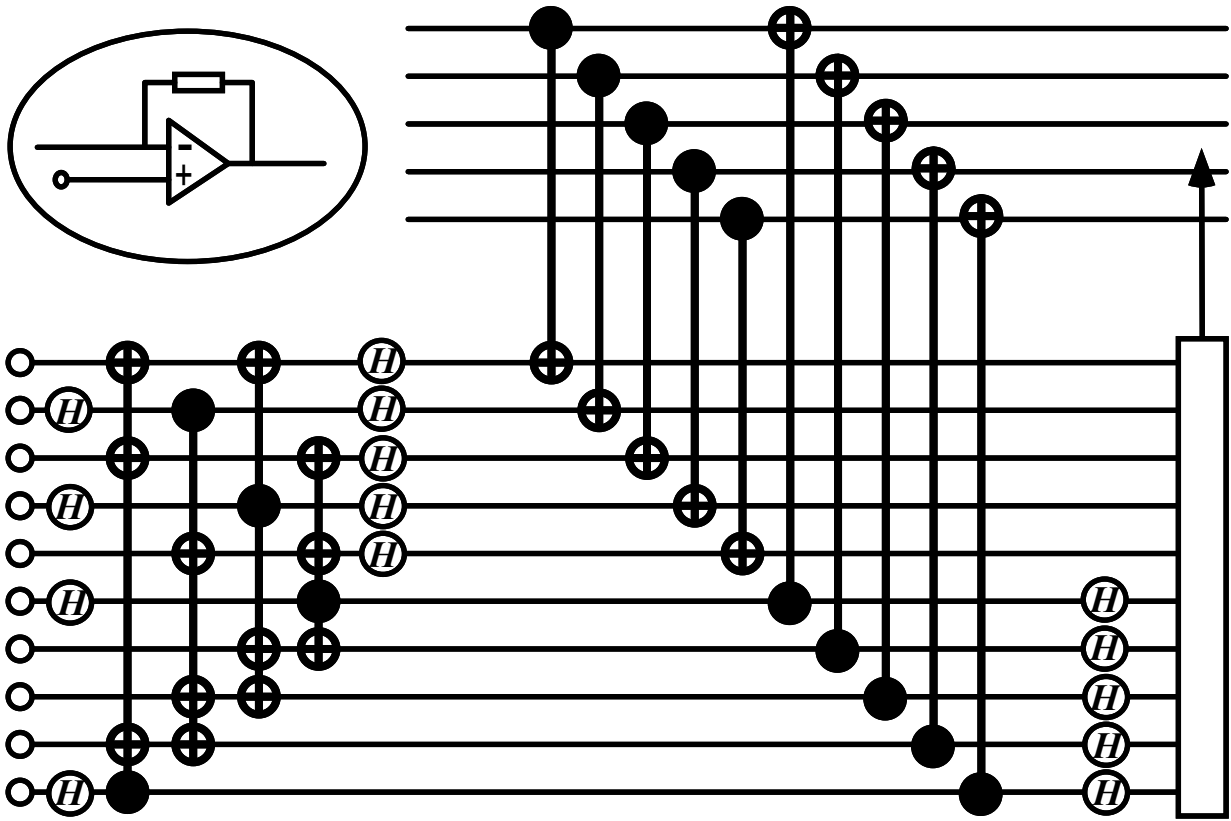
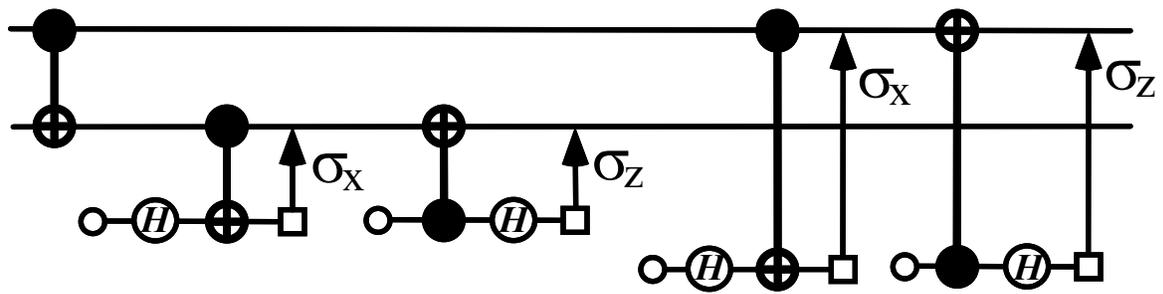
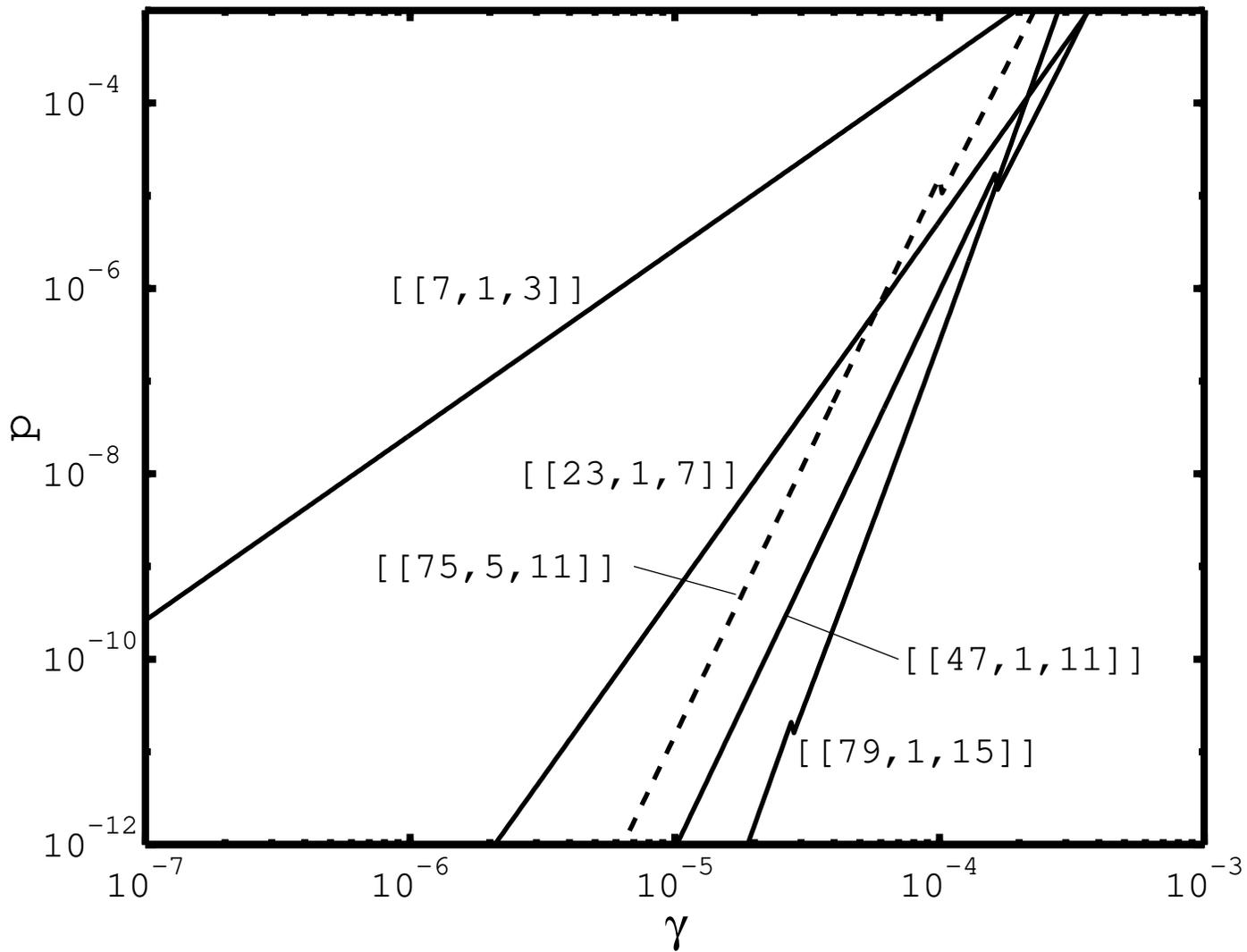
\end{document}